\documentclass[amsfonts,aps]{revtex4}

\begin{document}

\title{Comment on ``Sound Modes broadening in Quasicrystals'' and its Peer Review} 

\author{Gerrit Coddens}

\address{Laboratoire des Solides Irradi\'es, Ecole Polytechnique,\\
F-91128-Palaiseau CEDEX, France}   
\date{today}   


\widetext 
\begin{abstract}
Recently de Boissieu {\em et al.} 
proposed an explanation for
the broadening of acoustic modes observed in quasicrystals (QC).
It is the transcription  of a well-known model used for glasses.
We raise two fundamental objections against applying it 
to QC. After the text of the Comment we report
the methodology that has been used 
to thwart the publication of this Comment, which is
perfectly valid.
\end{abstract}


\maketitle
\narrowtext

{\section{The Comment}}

Recently, de Boissieu et al. \cite{deBoissieu} proposed a mechanism to explain
the broadening of phonons starting from a threshold
wavevector in QC, as is systematically
observed in experimental data. The mechanism 
(the coupling of sound waves to a heath wave) is quite general
and not new, as it has been known for many years in glasses.
There is obviously no doubt that the mechanism itself is sound but
we must take issue with the way it is
being used through a transcription to the field of quasicrystals.
Below, we explain our objections on two levels:

(1) On a general level, phonon broadening is an intrinsic 
property of quasicrystals,
even in a completely harmonic model, such that there
is {\em a priori} no need for the introduction of an assumption 
of anharmonicity in the form of a coupling.

(2) On a more detailed level, the authors try to blow new life into  
a cluster scenario proposed 
by Janot et al.\cite{Janot}, by using the idea of localized modes on clusters
 as a basic ingredient for the
microscropic realisation of the mechanism. The localized
modes on the clusters are this time no longer directly 
responsible for the broadening,
but they are proposed to be flat modes, that 
couple to the sound waves. To cite them verbatim:
"The building bricks of all QC structures are atomic clusters.
These clusters are not mere geometrical constructions
but real physical entities responsible for specific features
in the QC vibrational spectrum (e.g., responsible for localized modes)".\cite{remark}

Hence on two levels   
assumptions are introduced that are not granted or even not needed,
while through the presentation the reader might be left with the impression
that the experimental data present evidence in support of these
assumptions. 

(1) There exists an extensive
literature on phonons in QC, e.g. on the Fibonacci chain,
based on the transfer matrix method.\cite{transfer}
What transpires from such (rigorous) 
studies is that the phonon eigenmodes
are not at all periodic and even not quasiperiodic.
(There exist e.g.\\  

\vspace{1.5cm}

\noindent  so-called recurrent eigenmodes:
As a function of its position in space
the amplitude of a mode can  exhibit  humps around
the values $a \times \tau^{n}$, where $\tau$ is the golden mean,
$n$ is the set of integers, and $a$ a constant (length).
In between the humps, the amplitude of the mode is never
zero but it can be very weak).
The simple argument of the non-quasiperiodicity of the 
eigenmodes shows that
their Fourier analysis in terms of wave vectors Q will not show
dispersion curves of zero line-width as in crystals,
but broader features, and that this broadening
is not in energy (as de Boissieu et al. think)
but in the wave-vector Q. For the recurrent modes, specialists even wonder if 
it is a mathematically legitimate to take it for granted that they have
a Fourier transform in Q-space,
and in the case it is not, what kind of information the neutron data 
contain about these modes.

These features occur even in models that are perfectly harmonic
without any coupling between the modes.
Of course calculations in the harmonic approximation
on successive approximants can never reveal
such broadening, because they give rise
to zero linewidths by definition. A key intrinsic difficulty
of the problem is thus just being missed by such an approach, 
but it is from the blind spot
inherent to  such calculations that one might feel the urge to
inject additional assumptions of anharmonicity into the problem
under the form of localization or couplings. 

It would be cheap to push the present objections away, 
by arguing that real QCs are not one-dimensional, and that
surely there will be a loophole of escape from these objections,
when we go to higher dimensions.
Such vague arguments would (a) reverse the charge of proof,
and (b) contain a tacit denial of the horrendous difficulty of eigenvalue problems
(with the correct boundary conditions)
on quasiperiodic structures. We may add to this that (c) anharmonicity 
can be experimentally evidenced by
the temperature dependence of the Debye-Waller factor.\cite{Mossbauer}

When one drops full mathematical rigor, approximate eigenmodes 
that are periodic or quasiperiodic, leading to zero approximate line widths,
are  only expected to exist in the long-wavelength limit.
This correlation between absence of broadening and long wavelengths 
is confirmed by the experimental observations.
But the cluster mechanism proposed by de Boissieu et al.  only holds
in the limit in which acoustic
wavelengths are much larger than a typical cluster size, i.e.
the very limit
where on the basis of the preceding arguments  the broadening 
of the line widths is rather expected (and observed)
to be minor.

(2) In key positions of their paper (the abstract and the final conclusion)
the authors stress the important r\^ole they propose clusters to play
in the microscopic realisation of their mechanism. 
The authors discuss the relevance of {\em isolated} clusters,  
and present the issue
in terms of a unique isolated spherical inclusion in a vibrating medium
(e.g. a metallic sphere inside a rubber medium).
They dismiss such a scenario with great emphasis, but this 
does not really clarify 
the assumptions that underly their paper with respect to the issue
if there are isolated clusters in quasicrystals:
The scenario chosen to assess this issue is too obviously wrong,
and the real issue if there are  isolated clusters in quasicrystals
is more subtle. 

We have to address here a difficult situation of 
possibly ambiguous terminology, because
the authors do indeed introduce a concept of ``isolated'' clusters, 
different from the one that might be inferred from their presentation.
Let us call the type of isolation evoked by the model 
of a {\em unique} spherical inclusion 
{\em ``type 1''} and the type of isolated clusters
used by the authors {\em ``type 2''}.
The inadequacy of the type-2 isolated cluster model does
not hinge on the abundance of clusters in the structure as one might
infer from the type-1 model with its unique cluster.
It rather consists in tacitly denying
the importance of boundary conditions
in the set of coupled differential equations that
describe the phonon problem (and which are in general expressed
in terms of a dynamical matrix).

The point is easily understood as follows.
Take e.g. a one-dimensional crystal that is based
on the periodic repetition of the small motif $LSLLS$
taken from the Fibonacci chain.
The crystal is thus $...LSLLS.LSLLS.LSLLS...$.
Based on visual clues we could claim that  $LSLLS$ is a cluster,
and that the crystal is a dense packing of clusters.
The eigenmodes and eigenvalues of an isolated cluster $LSLLS$
are completely different from those
of the crystal $...LSLLS.LSLLS.LSLLS...$.
In fact, the cluster $LSLLS$ has  six discrete flat modes 
with a certain dynamical form factor,
$S({\bf{Q}})$ that extends throughout reciprocal space.
The modes are thus flat due to the finite extension of the cluster in space.
The modes of the crystal are completely different: They do not correspond
to a few isolated discrete energies, but build a whole dispersion curve
and for each eigenmode (i.e. each energy) in the acoustic regime the $Q$-dependence 
is Dirac-like in reciprocal space (if we limit ourselves to one
Brillouin zone).

Of course, it would be completely inappropriate to claim
on the basis of the visual clue that we can discern clusters
$LSLLS$ in the crystalline structure that there are flat modes in the crystal,
and that there would exist a coupling between sound waves and 
these flat modes in the crystal.
The flaw in such a reasoning is uniquely based
on a tacit change of the boundary conditions: It replaces periodic
boundary conditions (with a rather smooth variation of the force constants across
the ''cluster'' boundaries)
by an abrupt discontinuity at the ``surface'' of the imaginary cluster.

In other words: Type-1 isolation refers to the absence of similar clusters in the 
surroundings, the cluster is alone.
Type-2 isolation refers to a decoupling of the cluster from the surroundings
in terms of the force constants.
And of course what is relevant for the phonon problem is not if the cluster is alone
(type-1 isolation) but how the clusters couple to their
surroundings at their supposed boundaries (type-2 isolation).
Even in the example of a unique metallic sphere in a rubber medium,
it is type-2 isolation that is physically relevant.

QCs are not periodic, and are subject to other boundary conditions
than the ones that prevail in a crystal (see below).
But this certainly cannot mean that our example would not be 
appropriate and that the 
introduction of ``clusters'' by the authors
would have less of a hidden problem with the prevailing boundary conditions,
because the crucial point of our objection lies in the postulated abrupt discontinuity,
not in the periodicity.
When I am raising
an objection against the use of ({\em type 2}) isolated  clusters,
it can thus certainly not imply that I would have missed the passage in the paper
where the authors acknowledge that
there are no ({\em type 1}) isolated  clusters in quasicrystals.
The statement that there are no ({\em type 1}) isolated clusters in QCs 
(because they are a dense packing of clusters)
cannot hide the fact that it is the very use
of ({\em type 2}) isolated clusters  which  is the basic ingredient for the
microscropic interpretation proposed by the authors.

We could never warn the reader enough against the 
pitfall that would consist in getting one's attention
sidetracked towards the issue if there is  convincing evidence for
the presence of clusters in QCs or otherwise.
That would be certainly an interesting topic in its own right, 
but rather pointless
and misleading
in the present discussion, as
the issue if there are (type-2) {\em isolated} clusters in quasicrystals 
cannot be replaced by an issue if there are clusters in quasicrystals all together,
nor by the issue if clusters are physically meaningful in quasicrystals.
The verdict on the latter issues will moreover depend on the context
of the application: A possible cluster argument in a problem of stability
or electronic properties will be different from the one in a phonon problem.

The authors formulate the statement that clusters are not mere geometrical
constructions without any proof
as though it would be an obvious thruth, and the difficulty
that they can overlap is passed under silence. 
The claim that the origion of the flat modes observed in AlPdMn can be attributed to
 a localization on clusters is
also put forward without any proof.

To introduce the boundary conditions underlying their cluster assumptions, the
authors should have given arguments
that there is a discontinuity in the force constants
at the surface of these clusters.
In certain points on the cluster boundaries,
the contrary rather seems to be true,
viz. when instead of being isolated clusters 
overlap, which is often the case.
In such points it rather looks as though nothing in the whole set of the atomic 
forces between pairs of atoms
in a QC singles out a cluster as an isolated entity,
defined by such a discontinuity. The forces
between the atoms inside the clusters are not obviously different from those
between an atom of the cluster and a neighbouring atom that lies
just outside the cluster (but inside the overlapping cluster of the same type). 
A few phason jumps can create the illusion that a whole cluster has
jumped, which also clearly illustrates the relative arbitrariness
of assigning an atom to a cluster and  of suggesting
that a cluster would be an isolated entity whose existence would be
obviously defined by a discontinuity in the atomic forces at its surface. 

Mathematically spoken,
if a cluster is taken large enough it can even be a covering cluster
for the whole QC.  One can imagine a crystal that could be 
depictured as a (periodic) arrangement of physically acceptable, 
overlapping identical 
clusters of a certain size, and that would not lead to any localization
or broadening.
Any attempt to escape from this trivial objection 
must therefore forcedly end up in a discussion of the global, non-periodic 
arrangement of the clusters and their overlaps.

Discussing QC problems 
in terms of clusters rather than atoms, is thus just a kind of
renormalization procedure,
that merely shifts the intrinsic difficulty of 
non-periodicity to a different length scale, but does not tackle
the difficulty itself. 
It is a blunt denial of the subtlety and difficulty of the eigenvalue
problem to overemphasize the r\^ole of clusters.
We can illustrate this with the Fibonacci chain.
It starts with {\small $LSLLS.LSL.LSLLS.LSLLS.LSL.LSLLS.LSLLS.LLS. ...$},
where we have subdivided the sequence in building bricks $LSLLS$
and $LSL$.
Each of the ocurrences $.LSL.$ herein is seen to be followed
by $LS$ as both building bricks $.LSL.$ and $.LSLLS.$ begin with $LS$.
Hence the whole sequence can be seen as made from the ``covering cluster''
$LSLLS$, whereby we have to allow for overlaps $LS$, which appear
exactly at the positions where we have separated out $.LSL.$. 
Similarly we could even consider $LSL$ as a covering cluster (the overlap would
then be $L$).
Now, the phonon eigenmodes of the isolated sequences $LSL$ and $LSLLS$
can be calculated (from the corresponding $4\times 4$ and 
$6\times 6$ dynamical matrices). What does this handful of eigenmodes tell us about
the phonons of the Fibonacci chain? Hardly anything! As we pointed out above,
even the phonons of the periodic sequences based on 
$LSL$ or  $LSLLS$ do not give us the correct picture,
despite the fact that  in such sequences
the clusters are no longer completely isolated (which would be a completely irrealistic 
boundary condition) and  one at least allows 
for the point that they are {\em embedded} in a larger structure (which completely
changes the eigenvalue problem).

The idea of clusters $LSL$ and $LSLLS$ certainly has great eye appeal.
One might think at first sight that it must yield great insight
in the dynamics of the Fibonacci chain.
But as we explained above, all this is mere deception.
Already the overlap $LSLLSLLS$ of two clusters of the type $LSLLS$ will yield 
competely different solutions for the eigenvalue problem than $LSLLS$ itself.
The same basic objections about the boundary conditions 
 remain perfectly valid in the three-dimensional case,
such that the fact that we work on the one-dimensional case
does not present a loophole from these objections.
All the use of the clusters $LSL$ and $LSLLS$ allows us to do is to rewrite
the transfer matrix formalism in terms of matrices that
correspond to $LSL$ and $LSLLS$
rather than in terms of the more elementary matrices
that correspond to $S$ and $L$. 
This illustrates how replacing atoms by clusters is just 
a renormalization procedure, as we stated.
It is an underestimation of the complexity of eigenvalue problems
and their boundary conditions (which is {\em global})  to
suggest that they could be approached {\em locally} by focusing
one's attention to small building bricks. Putting the bricks together
just changes everything. 

At least in the present context we can thus state that
unless a rigorous  proof of the contrary is given,
it is  wise to adopt cautiously the conservative view point 
that the rigorous application of the idea of clusters, even if they 
look physically attractive,
has remained limited to just a convenient pictorial shorthand to
describe parts of the structure, nothing more.
We can appreciate from this discussion how both objections (1) and (2)
are linked, in the sense that both are based on a
tacit modification of highly sensitive details of an eigenvalue problem,
that is very hard to spot.
The example of how the recurrent modes completely escape the analysis
in terms of periodic approximants, shows to what kinds of catastrophies
such lack of rigor can lead.
Once again, this concern about rigor should not be misrepresented
by saying that I would claim that there are no clusters in quasicrystals,
or that clusters could not play a role in quasicrystals, etc...

Without any justification, 
the localized modes invoked are identified with the flat modes
that have been reported in AlPdMn, 
and a coupling 
mechanism between these localized modes and sound 
waves is proposed. We have two objections to
this: 

(a) Such an explanation for the flat modes
is just one between several other possibilities.
One of the alternatives is documented and can therefore not be ignored: 
By a scrutiny of the displacement patterns 
in their numerical simulations
Hafner and Krajci \cite{Hafner} were able to associate 
the flat modes with a restriction (``confinement'')
of the vibrations to disclination lines of 
atoms that are topologically
different from
average  (e.g. the atoms 
have a 13-fold coordination, rather than a normal 12-fold one).
This has nothing to do with the vibration on a cluster.

(b) The issue if the flat modes are due to a localization
on clusters is not open-ended within the present state of 
knowledge. It can be unambiguously settled.
It suffices to check if
the structure factor of a flat mode is indeed compatible
with the dynamical structure factor of a cluster vibration (as Buchenau
has done to prove his model for the dynamics of silica).
Although the dynamical structure factors of the
flat modes have not been published, it must be 
straightforward to extract 
this first-rank information from the authors' already 
existing data, and a numerical calculation of the vibrational 
spectrum of a Bergman or a Mackay 
cluster with realistic force constants, involving 
typically 33 to 55 atoms, is certainly not unfeasible.

Hence, before one can formulate any possible approach
of the type proposed by the authors,
it is a peremptory prerequiste that one first proves on the basis of existing data,
that (1) the observed structure factors of the flat modes are compatible
with an interpretation of these modes in terms of cluster phonons,
and (2)  that there are  anharmonicities
within the system, e.g.
on the basis of Debye-Waller factor anomalies
of which one has proved beyond any doubt
that they cannot possibly be attributed to an onset 
phason hopping. These are necessary but not sufficient,
minimal conditions that have to be met.
They stand completely free from any theoretical considerations,
 and therefore add up completely independently
to the two main objections  outlined in the present Comment.\\

{\section{The Peer Review of the Comment}}

We would like to give the reader an inkling about the methods that are used to have valid Comments
of this type rejected. I would have prefered to quote the referee reports literally rather
than paraphrasing them but it has been pointed out to me that it 
is illegal to reproduce referee reports
literally as this constitutes a violation of copyright.
I must state that I feel very uncomfortable about this. First of all, it kind of
diminishes my credibility and exposes me to cheap and easy criticism
that I am distorting the truth because I would not reproduce what has been written literally.
Secondly, when one literally reproduces what has been written,
one cannot be accused of being responsible for whatever that might be contained in it
and look disgracious,
while when one has to paraphrase it, one becomes subjectively accredited
with this responsibility.
Thirdly, it is apparently not enough that anonymous peer review
exposes people almost defenseless to the huge prejudices that can be inflicted
by sham peer review, especially when it becomes systematical
if some group has managed to completely
invade the horizon of an editorial board. Victims are this way also obstructed
from denouncing what they have undergone and making the community aware of it.
I must ask the reader to consider  how this obligation to mention
what has been stated only indirectly,
can only result in 
a down-sized and filtered account
of the adversity and the personal attacks I have been faced with. 

In a first reply de Boissieu stated that the verbatim quotation 
I made at the beginning of my Comment:
"The building bricks of all QC structures are atomic clusters.
These clusters are not mere geometrical constructions
but real physical entities responsible for specific features
in the QC vibrational spectrum (e.g., responsible for localized modes)"
would not be in his paper, and that my Comment would distort the point of view
of his paper by making quotations out of context.

He also stated that it was not appropriate to take issue with
the fact that his analysis is based on isolated clusters,
while on p.5 of his paper he had clearly stated that the picture of one isolated
cluster is not adequate for QCs, and that one should rather think of a QC
in terms of a dense packing of clusters.
The reader should not get confused by this would-be catching me in my own
contradictions. The contradictions are entirely from the hand of de Boissieu et al. themselves.
Yes, de Boissieu et al. state 
with great emphasis in the beginning of their paper that they
do not use isolated clusters, refering to the example of a metallic sphere
in a rubber medium. But, no, this
cannot hide that the whole argument
is exactly based on an assumed isolation of the clusters.
It is just that the isolation at stake is very different from the
one suggested by the example of a single metallic sphere in a rubber medium.
It is not by sorting two entirely different 
situations under a same descriptive phrasing
that they would become equal.
In fact, the problem of isolated clusters
is not one of numbers (``one'' against ``a dense packing''), as one
could infer from this reply, but one of mutual overlap.
In order to overcome the confusion that could result from this reply,
I introduced the definitions and the distinction between type-1 and type-2
isolation in my Comment.

de Boissieu further stated that the scientific content
of my comment would be very small, that it contained vague and general
views, that were in lack of scientific evidence and that
were not supported by recent papers. He stated that I overinterpreted the literature.

Finally, he replaced the true issue of my Comment, viz. that the clusters are not
isolated, by another issue, viz. if there are clusters in quasicrystals at all,
and then gave a detailed reply on this replacement issue, with several citations.
Even if this reply had been entirely correct, it did not address the issues
I had raised in my Comment.

With respect to my argument that it is not true that clusters are isolated in the sense
of having significantly stronger intra-cluster bonding,
de Boissieu stated that he certainly agreed that the existence of clusters
is still a matter of debate, but that it has not been proved
that the forces are of equal strength throughout the QC and that therefore my argument
had no firm ground. 
He added that the problem of the energetics
of QCs is very complicated and one could not expect the final solutions
to be given soon.

With respect to my citation of the work of Krajci and Hafner he stated that
this was wrong, because the five flat modes reported by these authors have an insignificant
participation ratio, and because there has been no analysis of their vibration patterns, 
such that their true nature is still not clear,
even if it is clear that they are associated with disinclination lines.

It was also stated that electron density measurements on a cristalline
approximant of AlReSi indicate a larger bonding character within clusters
than in between them. (This argument is clearly not general as can immediately
be appreciated from the fact that clusters very often overlap, which is the
real issue of my Comment).

It was also stated that Gratias approved the cluster approach,
while it very clearly transpires from Gratias' papers that he rather
very cautiously considers them as a convenient shorthand to describe the structure.

It suffices to point out that all this does not address at all
the issue I raised, which is that the clusters often overlap.
The point is thus not
if there are clusters at all in quasicrystals, but that these clusters are not type-2 isolated.

After I had replied to this, the correspondence was sent to two referees.
One of them appeared biased to me. But
eventually, both of them recommended publication with small
modifications. However, I learned later on that one of them wrote a seperate note to the editors
wherein he/she stated that he/she did not wish to review my manuscript any further,
because the manuscript would not be presented in the proper manner.
(The ``he/she'' reflect a terminology that was used by the APS).

When I had adapted my version to these comments, 
the editors of PRB requested that I should remove
the sentence from my paper that very explicitly stated
that one should be careful in not being sidetracked
towards the fake issue if there are clusters in QCs,
as the true issue is if these clusters are isolated in a very
specific sense. I had made a citation towards de Boissieu's 
reply to the editors in this respect. It was argued that
I would not have the right to cite correspondence from the peer review
process. It was also stated that the editorial board
was {\em ``positively inclined''} to accept my Comment for publication.
Judging that the way de Boissieu changed the issues was apt to mislead
a many reader, I reformulated the sentence such as to keep
its contents but to  remove the citation.
Then PRB put my paper ``on hold'' for six months, refusing to give any explanation.
They had done that already a first time in the review process. 
Perhaps I should have understood from this that my Comment was
not well considered by
the APS itself, and that this gentle use of force 
was meant to be discouraging enough to make me just give up.
Finally after nine months, they sent me all at once a report from a fourth referee.
There had very obviously not been the slightest reason to ask advice from a fourth referee.

This referee replaced again the issue if the clusters are isolated
by the false issue if there are clusters all together.
On this replacement issue he stated that an intense debate
was going on within the community, and  again developed a whole argumentation
about this non-issue, citing the model of closed electronic shells by Janot et al.,
the fracture experiments by Ebert et al., confirmed by numerical simulations by Roesch et al. 
and the cluster friction model by Feuerbacher et al.
It was also stated that the whole development based on the Fibonacci chain
was {\em insignificant}, and unsuitable to disprove
the assumption that there are clusters in QCs,
because real three-dimensional clusters are very different from
sections of the Fibonacci chain (e.g. in containing many shells).
That the looks of one- and three-dimensional clusters are 
different may very well be true, but is irrelevant for the real issue,
which is not if there are clusters in QCs, but that these clusters are not separated.
And the latter issue can equally well be explained on the Fibonacci chain as
on a three-dimensional model.
It was also stated
that as long there was no rigorous proof that there are no clusters
in QCs, it would be allowed to assume that clusters are present
and to build models on this assumption. Once again, the issue
is not if there are clusters all together, but that these clusters
are not isolated.

The referee also paraphrased me by stating
that I would have claimed that the considerations of de Boissieu are redundant,
because phonon broadening is an intrinsic property of QCs. He argued
that this might be plausible, but that it would be too simplistic,
because the broadening cannot be calculated, while de Boissieu et al.
would convincingly explain the experimental data.

I responded more or less as follows (I am forced to
paraphrase the exact statements from the report in order not to violate copyright):

I clearly wrote in my Comment that
"We could never warn the reader enough against the 
pitfall that would consist in getting one's attention
sidetracked towards the issue if there is  convincing evidence for
the presence of clusters in QCs or otherwise."
This sentence summarizes a number of arguments that are clearly developed
in my paper: The issue is not if there are clusters or otherwise.
The issue is that de Boissieu et al. ignore the  consequences
of the crucial fact that the clusters often overlap, and that
they therefore are not isolated (in the  type-2 sense defined in my paper).
(And the issue that the clusters are not isolated
is a very different one from the one that de Boissieu dismissed
after evoking an isolated metallic sphere in a rubber medium, which of the type-1
defined in my paper). The fact that I define two types of isolation
already clearly shows that the issue is not if there are clusters all
together in QCs. The issue is that these clusters are not isolated
in the type-2 sense.
I have a whole discussion of this in terms of boundary conditions
within the paper.
The referee finds it convenient to ignore this all together,
and cites work of Janot et al., Ebert et al., Roesch et al. and Feuerbacher et al.
as proofs for the existence of clusters while this is totally pointless,
as clearly explained in my paper and the sentence I quoted from it above.  
The referee thus goes resolutely for the pitfall
I warned against in my paper, which I formulated because de Boissieu
had attempted this elusive move already in his first reply.

The referee builds further on this swap towards a false issue, by stating that 
the example of the Fibonacci chain, which would take a
disproportionate large part of my Comment
could not be used to prove the cluster assumption wrong.
In my paper, the Fibonacci chain is not being used in order to prove
that there would be no clusters in a QC. As I already pointed out above,
the issue if there are clusters or otherwise is pointless
for our discussion. What matters is that the clusters
often overlap (i.e. are not isolated in the type-2 sense)
and in order to point this out, the Fibonacci chain
is as good as a full-fledged 3D model.
And the referee further insists on focusing the 
attention to this pointless issue
of the existence of clusters or otherwise when he writes
that three-dimensional clusters
 are ot comparable to sections of the Fibonacci chain,
 and that  it is legitimate to use the cluster assumption
 as long as there is no striking argument against it.

The latter contains actually a reversal of de Boissieu's charge of proof:
One could make as many unproved claims as one likes,
it is up to others to prove them wrong. de Boissieu and the referee
know very well that this is not viable. This shines through clearly enough
when it comes down to attacking my work, rather than defending
the work of de Boissieu, and one tries to make prevail
totally irrelevant objections against it to make us wonder if  perhaps 
I did not fail to meet my charge of proof on some very tiny loopholes.

One of these irrelevant objections is that the Fibonacci chain would
not be pertinent in the present discussion.
The arguments developed on the Fibonacci chain
are not affected by his irrelevant distinctions between the Fibonacci chain
and 3D QCs he would like us to believe crucial.
The Fibonacci chain is perfectly suited for discussing the 
type-2 isolation and other issues at stake.
Making the same points on a 2D or 3D QC would require to
include into the paper elaborate Figures to show the clusters 
and how they overlap, etc...
While with the Fibonacci chain one can describe the whole situation
exhaustively and very clearly by referring
to the letters L and S. One issue of my paper is that
the possible overlap of the clusters (the lack of type-2 isolation) is pointing out
a tacit cheat about the values of the interatomic forces: It tacitly implies that
the bonds between atoms within the cluster are stronger than the bonds between atoms
of a cluster and surrounding atoms. That this is not true is totally obvious
when two clusters A and B overlap, as discussed in my paper.
And the referee can check it also on a 3D model. It is nothing specific
for the Fibonacci chain only.
This entails that the clusters are not at all doing what
de Boissieu claims they are doing.
This point is exactly the same one as voiced by Henley
in his paper "Clusters, phason elasticity, and entropic
stabilization: a theory perspective" 
(Phil. Mag. 86, 1123 (2006): Ames conference proceedings) in the first sentence
of the section "2. Clusters" on page 1124.
It is this issue, and not the mere presence of clusters or otherwise, that is essential
and makes de Boissieu's position untenable.
To make this untenable position prevail nevertheless,
the referee carefully eludes discussing this crucial point of overlap.
He diverts the attention away from it
by hammering incessantly on the irrelevant issue
if clusters exist all together.

I also addressed the paraphrasing of my argument that
the use of clusters would be redundant. 
In fact, this is a completely false presentation of the issues, as anyone who reads the
paper carefully can see. 
The referee operates a very subtle shift
when he paraphrases my objection by stating that I would have claimed 
the considerations of de Boissieu et al. are redundant.
As far as I can see I wrote:
"There
is {\em a priori} no need for the introduction of an assumption 
of anharmonicity in the form of a coupling."
If one thinks carefdully about it, this does not mean that
the considerations {\em are} redundant, but that they {\em could be}
redundant. The snag to this almost subliminal shift is that if I had claimed that
the consideratrions {\em are} redundant I would be invested with a charge to prove it.
While if the considerations {\em could be} redundant, it is the charge of proof of
de Boissieu that has not been properly met. Hence this is 
a  hidden reversal of the charge of proof, that is very hard to spot.
That phonon broadening is an intrinsic property
is not merely plausible as he states, it is a mathematical certainty,
because the eigenmodes are not quasiperiodic.
Hence here the referee unduly questions (again in an inoffensive looking way)
 an established obvious factual mathematical truth.
That nobody is able to solve the horrendously difficult problem
of the  calculation of the q-dependence of the intrinsic broadening,
does not exclude that the broadening observed could be entirely due
to this intrinsic broadening. Let one please  not jump
onto this sentence to put again things in my mouth that I do not say.
I do not say that the broadening {\em is} entirely intrinsic,
I say that  cannot be excluded that it could be entirely intrinsic.
The mechanism of intrinsic broadening has at least 
the merrit that it is physically sound, while
the cluster scenario is conclusively proved wrong by e.g. the 
type-2 isolation issue, which the referee carefully eludes to discuss. 

It is therefore ridiculous to exploit
the difficulty of the problem of calculating the intrinsic broadening
to compare this scenario unfavourably with the (illusory) succes of the cluster model
by stating that the result of the $Q$-dependence of the line width
presented by Boissieu et al. is convincing. 
 The model is proved wrong and
that the data can be fitted with a polynomial of the fourth degree
is hardly informative and a finding that could be derived
from scores of other models. 

I have never stated that the broadening would be uniquely intrinsic.
As I explained it already above
I have only evoked this as a possibility.
Because, what the argument of the intrinsic broadening
was intended to show is that the assumptions de Boissieu makes are 
 in lack of justification, by giving a counter example. 
E.g. the tacitly implied assumption that there is anharmonicity
is gratuitous. In view of the possibility
of intrinsic broadening which will occur even in a completely harmonic model,
it is  a peremptory prerequiste to prove that the forces are anharmonic before
one can introduce the assumptions that underly de Boissieu's model. 
 Asking to develop the intrinsic broadening scenario 
into a full calculation
as the referee does is, again, a reversal of the charge of proof.
Moreover it tries to saddle me up with the obligation
of a demonstratio diabolica.

The editorial board of PRB refused to consider my reply to this referee report.
They even refused to state this refusal.
They just moved on towards a statement that this ended the review of my Comment.
They eluded answering by addressing non-scientific issues, 
maintained my Comment
rejected and even eluded responding to an appeal
of mine. They had artificially made the whole procedure last for more than two years.
They even recommended that I try to have it published in another journal,
because other journals have other criteria for approval than the APS.

We may finally point out that we already had attempted to write a Comment on the artificial
cluster issue, back in 1993 when it was introduced in reference \cite{Janot}.
Janot et al. dropped an off-hand comment on my work towards the end of that
paper that my interpretation of the quasielastic data in terms of phason hopping
would be wrong. In reality, their data did not warrant such questioning
of my work. In fact, reference \cite{Janot} reported a failure to observe
the quasielastic scattering that I had measured and that corresponded to 
the decrease of the elastic intensity
when the temperature was raised. Such elastic data can never be used
to challenge the interpretation of the much more detailed and specific
quasielastic data. Nevertheless, Janot et al. did this,
denigrating my work. I had to discover this as an accomplished
fact in the published literature. To undo the damage, I was forced to write a Comment 
to reference \cite{Janot}, with reversed rights of reply.
Using this reversed situation,
Janot answered that my samples were suspicious and that the quasielastic signal
was due to preferential segregation of Cu into the grain boundaries.
On the editorial board of Physical Review Letters S. Moss stated that he
felt much more sympathetic towards my arguments, but that the exchange would be too long to 
publish it in Physical Review Letters. S. Moss and R. Schuhmann suggested 
to me that I send it to Physical Review B.
But when I did this, I was told that Physical Review B does not handle
Comments on papers published in Physical Review Letters.
In my Comment I had pointed out that the cluster scenario
was analogous to scenarios used in glasses.
But in reference \cite{deBoissieu} it is stated towards the end,
that after the work was finished, the authors discovered
that a similar approach had been used in glasses.

After the rejection of my Comment on \cite{Janot}, a proposal of mine for beam time
at the ILL to measure phason dynamics was rejected on the basis a statement by Dubois in the 
scientific evaluation committee
that the experiment had already been done by Janot. It was just not true.
When I protested, and I expressed my fears
that my ideas would be stolen, Janot wrote a letter to me 
with copy to the director of the ILL,
 wherein he stated
that I would be paranoid, and that they did not intend to measure
phason dynamics. A few months later he and de Boissieu 
made the experiment in my place on IN16, but they melted their sample.
They had made their attempt to measure phason dynamics with the same type of sample,
on the same type of instrument, in the very same Q-range,
with the same energy resolution, and in the same energy and temperature range.
Nevertheless, they wrote an ILL report about it wherein they stated
that this experiment would be different from mine and wherein they 
reported that {\em they} had figured out in the meantime that the interpretation 
of the Debye-Waller factor 
in \cite{Janot} was wrong. 
The interpretation of the temperature dependence of the Debye-Waller factor had been the only
element of justification on which the whole introduction of the cluster issue 
had been based. It was wrong.
And already at that stage, the obvious error in the reasoning
had been that the clusters are not isolated but overlap.
Nevertheless, these issues were introduced again in reference \cite{deBoissieu}.


\begin{thebibliography}{9}
\bibitem{remark} (a) Our verbatim citation can be found on page 5, 
at the top of the second 
column of reference [1].
(b) The authors qualify the optic modes observed
in QCs as ``dispersionless''. As acoustic phonons in the long-wavelength limit
are also dispersionless,
we prefer the  terminology ``flat'',
which not only indicates that the slope is constant with 
the wave vector ${\bf{Q}}$, but also
that this constant is zero.
\end{thebibliography}
\end{document}